\documentclass[aps,prb,twocolumn,superscriptaddress,showpacs]{revtex4}
\usepackage{epsfig}
\usepackage{amsmath}
\usepackage[dvipsnames,usenames]{color}

\begin{document}

\title{Interplay of localized and itinerant behavior in the one-dimensional Kondo-Heisenberg model}

\author{ Neng Xie}
\affiliation{Beijing National Laboratory for Condensed Matter Physics and\\
Institute of Physics, Chinese Academy of Sciences, Beijing 100190, China}
\author{ Yi-feng Yang }
\email[]{yifeng@iphy.ac.cn}
\affiliation{Beijing National Laboratory for Condensed Matter Physics and\\
Institute of Physics, Chinese Academy of Sciences, Beijing 100190, China}
\affiliation{Collaborative Innovation Center of Quantum Matter, Beijing 100190, China}

\date{\today}

\begin{abstract}
We use the density matrix renormalization group method to study the interplay of the localized and itinerant behaviors in the one-dimensional Kondo-Heisenberg model. We find signatures of simultaneously localized and itinerant behaviors of the local spins and attribute this duality to their simultaneous entanglement within the spin chain and with conduction electrons due to incomplete hybridization. We propose a microscopic definition of the hybridization parameter that measures this "partial" itinerancy. Our results provide a microscopic support for the dual nature of $f$-electrons and the resulting two-fluid behavior widely observed in heavy electron materials.
\end{abstract}
\maketitle

Coexistence of superconductivity with competing magnetic orders has been observed in many heavy electron superconductors \cite{Park2006}. This exotic phenomenon supports the dual (simultaneously itinerant and localized) nature of $f$-electrons. A phenomenological formulation of this idea in a two-fluid model has provided a unified explanation for a variety of anomalous properties of heavy electron materials and yielded a number of surprising predictions including the universal logarithmic temperature scaling of the heavy electron density of states \cite{Nakatsuji2004,Curro2004,Yang2008,Yang2008b,Yang2012,Yang2014}. The two-fluid model provides a possible solution to the Kondo lattice problem and a simple framework for understanding the heavy electron physics \cite{Yang2012}. It proposes two coexisting and competing quantum fluids in the normal state of heavy electron materials: a spin liquid of partially hybridized $f$-moments and a heavy electron liquid that emerges as a composite state of conduction electrons and magnetic fluctuations of the local moments due to collective hybridization. Similar two-fluid behavior has also been observed in the cuprate \cite{Barzykin2009} and pnictide \cite{Dai2012} superconductors. However, despite much effort \cite{Barzykin2006,Zhu2011,Choi2012,Jiang2014}, a satisfactory microscopic theory of the two-fluid behavior has not been achieved. In particular, it is not clear what the duality exactly means microscopically, how the $f$-electrons can be simultaneously itinerant and localized and how one can measure this "partial" itinerancy.

In this work, we study the interplay of the localized and itinerant behaviors in the one-dimensional (1D) Kondo-Heisenberg model using the density matrix renormalization group (DMRG) method \cite{White-92,White1993,Schollwock2005}. The Kondo-Heisenberg model contains by definition two distinct components: the conduction electrons and the local spins. The DMRG method allows us to numerically calculate the momentum distribution of the conduction electrons and the correlation spectrum of the local spins and use these to track the detailed evolution of both components with varying Kondo coupling. A joint analysis of these quantities suggests that each local spin entangles simultaneously with other local spins and conduction electrons in the intermediate coupling regime, giving rise to signatures of emergent heavy electrons in a background of partially hybridized spins. This provides a natural basis for the dual behavior of $f$-electrons and the two-fluid physics observed in heavy electron materials.

The Hamiltonian of the 1D Kondo-Heisenberg model may be written as
\begin{align}
H= & -t\sum_{n=1}^{N-1}\sum_{\sigma=\pm}\left ( c_{n,\sigma}^{\dagger }c_{n+1,\sigma}+{\text H.c.} \right )
\nonumber \\
  & +J_{K}\sum_{n=1}^N\vec{s}_{n}\cdot\vec{S}_{n}+J_{H}\sum_{n=1}^{N-1}\vec{S}_{n}\cdot\vec{S}_{n+1},
\label{hamiltonian}
\end{align}
where $t$ is the hopping parameter of the conduction electrons, $J_{K}>0$ is the on-site antiferromagnetic Kondo coupling between the conduction electrons and the local spins, and $J_H>0$ is the nearest-neighbor Heisenberg exchange coupling within the spin chain. $\vec{S}_{n}$ is the spin operators of the local spins, $c_{n,\sigma}^{\dagger }$($c_{n,\sigma}$) creates (annihilates) a conduction electron of spin $\sigma$ at the $n$-th site, and
$\vec{s}_{n}=\sum_{\alpha,\beta}c_{n,\alpha}^{\dagger }(\vec{\sigma}/2)_{\alpha,\beta}c_{n,\beta}$, where $\vec{\sigma}$ is the Pauli matrices, is its spin operator. Away from half-filling, the local spins could be ferromagnetically correlated for $J_H=0$ \cite{Tsunetsugu1997,Gulacsi2004}. A finite $J_H/t=0.5$ is introduced to suppress the ferromagnetic correlations \cite{Moukouri1996}. For DMRG calculations, we use the DMRG++ code \cite{Alvarez2009} with open boundary conditions and keep $500$ block states for calculations on a lattice of $N=50$ sites. The results are verified with different lattice sites and block states and found to be converged. The good quantum numbers are the average occupation number of the conduction electrons, $n^c=N^{-1}\sum_{n\sigma}c_{n\sigma}^\dagger c_{n\sigma}$, and the $z$ component of the total spin, $S^z_{tot}=\sum_{n}(s_n^z+S_n^z)$, which is zero in the ground state for $J_H/t=0.5$.

\begin{figure}[t]
\begin{center}
\includegraphics[width=0.4\textwidth]{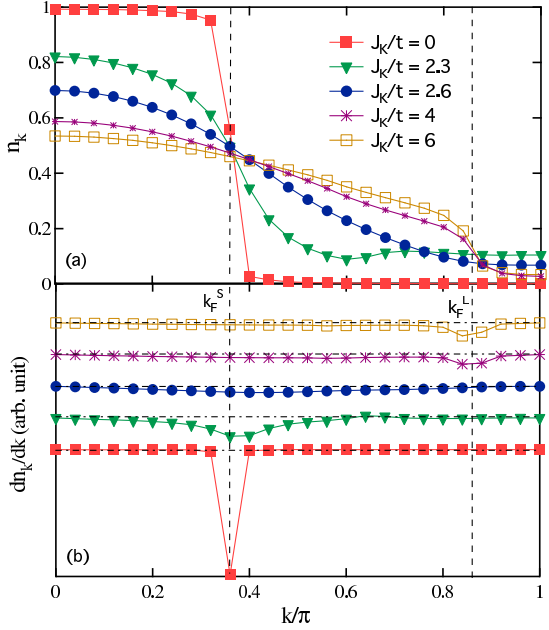}
\caption{(color online)  (a) The momentum distribution, $n_k$, of the conduction electrons and (b) The derivative, $dn_k/dk$, for $n^c=0.72$ and $J_{H}/t=0.5$. The dashed lines mark the small and large Fermi wave vectors at $k_{F}^{S}=0.36\pi$ and $k_{F}^{L}=0.86\pi$, respectively. }
\label{Fig:nk}
\end{center}
\end{figure}

\begin{figure}[t]
\begin{center}
\includegraphics[width=0.406\textwidth]{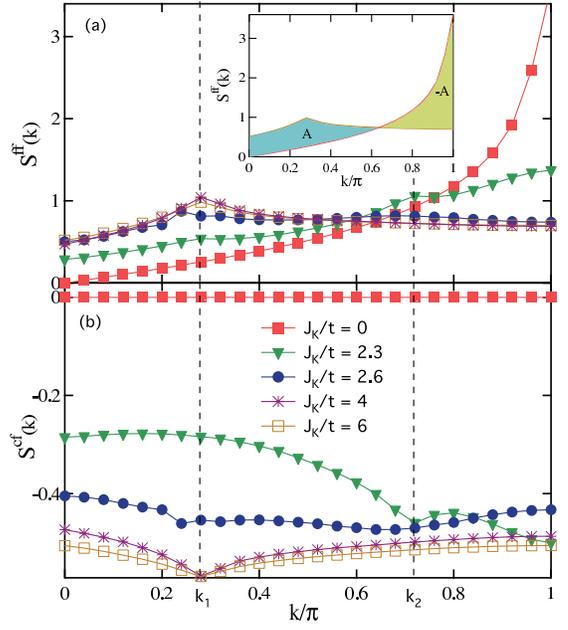}
\caption{(color online)  (a) The spin-correlation spectra of the local spins and (b) the hybridization spectra for varying $J_K$. The parameters are $n^c=0.72$ and $J_{H}/t=0.5$ and the inset illustrates the difference of $S^{ff}$ at $J_K=0$ and a finite $J_K$, $A(J_K$), used in the calculation of $f_A(J_K)$. }
\label{Fig:Sff}
\end{center}
\end{figure}

We first study the momentum distribution function of the conduction electrons,
\begin{equation}
n_k=\frac{1}{N}\sum_{n,m,\sigma}e^{ik(n-m)}\langle c_{n,\sigma}^{\dagger}c_{m,\sigma}\rangle.
\end{equation}
Fig.~\ref{Fig:nk}(a) shows the variation of $n_k$ with increasing $J_K$ for $n^c=0.72$ and $J_H/t=0.5$. For small $J_K$, $n_k$ changes rapidly from unity (fully occupied) at the center of the Brillouin zone to zero (unoccupied) at the zone edge. We calculate the derivative, $dn_k/dk$, and, as may be seen in Fig.~\ref{Fig:nk}(b), for small $J_K$ the most rapid change occurs at $k_F^S=\frac{\pi}{2}n^c=0.36\pi$, which corresponds to the small Fermi surface of free conduction electrons. At $J_K/t=2.3$ and 2.6, the distribution is strongly modified due to the coupling to the local spins; the maximal slope in $n_k$ is suppressed but always occurs at $k_F^S$ until it suddenly shifts to a different wave vector, $k_{F}^{L}=0.86\pi$, for $J_K/t>2.6$, as is plotted in Fig.~\ref{Fig:nk}(b). A simple calculation shows that $k_F^L=(1+n^c)\pi/2$, corresponding to the so-called large Fermi surface that has incorporated in it the local spins. Further increasing $J_K$ has little effect on the overall structure of $n_k$, indicating that the whole system has reached a strong coupling limit with a well-defined large Fermi surface. This limit may be easily understood: each local spin needs one conduction electron to form a local Kondo singlet so that effectively there are $1-n^c$ holes per site moving in the sea of the singlet background; the Fermi wave vector of these mobile holes is, $k_F^h=\pi-\pi(1-n^c)/2=k_F^L$, exactly at the large Fermi surface of the heavy electrons. The situation is more complicated in the intermediate regime around $J_K/t=2.6$, where $n_k$ changes smoothly and the conduction electrons spread all over the Brillouin zone so that there is no well-defined Fermi surface. In this regime, the conduction electrons and the local spins are strongly hybridized but not yet bound together to form local spin singlets. Their hybridization is collective and highly nonlocal.

Accompanying the gradual redistribution of the conduction electrons in the Brillouin zone with increasing $J_K$ is the change of the spin-correlation spectrum in the spin chain,
\begin{equation}
S^{ff}(k)=\frac{1}{N}\sum_{n,m}e^{ik(n-m)}\langle \vec{S}_{n}\cdot\vec{S}_{m} \rangle.
\end{equation}
The calculated spin-correlation spectra are plotted in Fig.~\ref{Fig:Sff}(a). At $J_K=0$, the local spins are decoupled from conduction electrons and form themselves a global spin singlet; we find a sharp peak at $k_0=\pi$ from their antiferromagnetic nearest-neighbor coupling, as predicted for the 1D Heisenberg model in the continuous limit \cite{Mikeska2004}. Increasing $J_K$ gradually suppresses the peak at $k_0$, indicating that the local spins are getting hybridized. At $J_K/t=2.3$, two additional peaks emerge at $k_1=0.28\pi$ and $k_2=0.72\pi$, but further increasing $J_K$ suppresses the peak at $k_2$. Only the peak at $k_1$ keeps increasing and eventually develops into a well-defined cusp. For $J_K/t>4$, the spin-correlation spectrum remains essentially unchanged and resembles that of a free electron system. Similar structures are also seen in the hybridization spectrum,
\begin{equation}
S^{cf}(k)=\frac{1}{N}\sum_{n,m}e^{ik(n-m)}\langle \vec{s}_{n}\cdot\vec{S}_{m} \rangle,
\end{equation}
as plotted in Fig.~\ref{Fig:Sff}(b). We find $k_2=2k_F^S=0.72\pi$, which might originate from the induced effective exchange coupling by the conduction electrons around the small Fermi surface, while the peak at $k_1=2\pi-2k_F^L=0.28\pi$ is associated with the formation of the large Fermi surface at $k_F^L$. 

\begin{figure}[t]
\begin{center}
\includegraphics[width=0.45\textwidth]{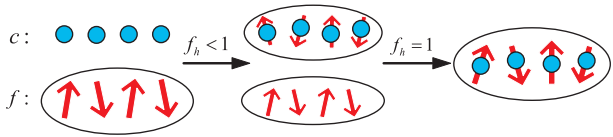}
\caption{(color online) An illustration of the two-fluid scenario. The local spins are only partially entangled with conduction electrons for the hybridization parameter, $f_h<1$, in the weak and intermediate coupling regimes.}
\label{Fig:2fluid}
\end{center}
\end{figure}

Combining these observations from $n_k$ and $S^{ff}(k)$ suggests that heavy electrons are a composition of conduction electrons and the local spins. The gradual change in the shape of $S^{ff}(k)$ and $S^{cf}(k)$ with increasing $J_K$ reflects the suppression of antiferromagnetic correlations within the spin chain and the corresponding emergence of composite heavy electrons around the large Fermi surface. In addition, the fact that all three peaks show up in the spectrum at around $J_K/t=2.3$ suggests the emergence of heavy electrons in the background of partially hybridized local spins, which are entangled at the same time with conduction electrons and other spins in the chain in order to fully dissipate their magnetic entropy in this intermediate regime.  Macroscopically, these may be seen in thermodynamic or magnetic measurements as coexisting itinerant heavy electrons and residual localized spins in the chain, as illustrated in Fig.~\ref{Fig:2fluid}, leading to the observed dual behavior of  $f$-electrons in heavy electron materials.

We note that the spins in the Kondo-Heisenberg model are by definition always local even though they may appear to be itinerant through collective hybridization with conduction electrons. Their spin-correlation spectra obey the following sum rule \cite{Mikeska2004},
\begin{equation}
\int_0^\pi \frac{dk}{\pi}\,S^{ff}(k)=\frac{1}{N}\sum_{n}\langle \vec{S}_{n}^2 \rangle=\frac34.
\end{equation}
The total spectral weight is conserved independent of the Kondo coupling. We can therefore use the spectral weight of the emergent structures as a measure of their localized or itinerant fraction. However, as seen in Fig.~\ref{Fig:Sff}(a), the two components at $k_0$ and $k_1$ are strongly mixed and hard to separate in the momentum space. 

\begin{figure}[t]
\begin{center}
\includegraphics[width=0.45\textwidth]{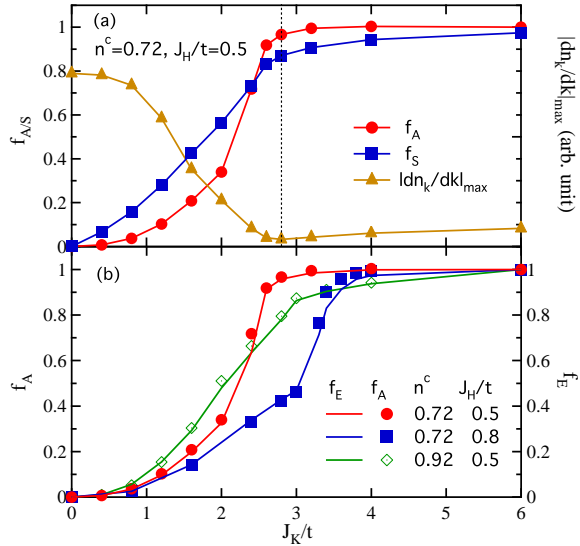}
\caption{(color online) (a) Comparison of the proposed hybridization parameter, $f_A$, the average local hybridization, $f_S$, and the maximal $|dn_k/dk|$ shown in Fig.~\ref{Fig:nk}(b) as a function of $J_K$. (b) Comparison of $f_A$ with the normalized antiferromagnetic correlation energy, $f_E$, as a function of $J_K$.}
\label{Fig:f}
\end{center}
\end{figure}

To proceed, we note that the local spins form a spin liquid at $J_K=0$, whereas they behave totally as itinerant heavy electrons at large $J_K$. Because of the sum rule, the conversion from the spin liquid state to the emergent heavy electrons with increasing $J_K$ may be seen as a gradual deformation of the spectrum that continuously connects these two limits. We may propose
\begin{equation}
f_A(J_K)=\frac{A(J_K)}{A(J_K/t=6)},
\end{equation}
where $A(J_{K})$ denotes the spectral weight transfer from $k_0=\pi$ to the emergent structure around $k_1=0.28\pi$ and is approximately given by the area enclosed by the two curves of $S^{ff}(k)$ at the chosen $J_{K}$ and $J_{K}=0$, as is illustrated in the inset of Fig.~\ref{Fig:Sff}(a). Although $f_A$ calculated in this way may not be the exact spectral weight of the emergent heavy electrons, we believe it is a good approximation to start with. For comparison, we also calculate the local hybridization, $\langle \vec{s}_n\cdot\vec{S}_n\rangle$, and define,
\begin{equation}
f_S(J_K)=\frac{N^{-1}\sum_{n}\langle \vec{s}_n\cdot\vec{S}_n\rangle_{J_K}}{-3n^c/4},
\end{equation}
where $-3n^c/4$ is the average local hybridization in the strong coupling limit. The results for $f_A$ and $f_S$ are plotted in Fig.~\ref{Fig:f}(a) for $n^c=0.72$ and $J_H/t=0.5$. Both quantities increase with increasing $J_K$ and show qualitative agreement. Their difference may be understood from the strong $k$-dependence of $S^{cf}(k)$. As plotted in Fig.~\ref{Fig:Sff}(b), with increasing $J_K$, the strongest hybridization changes from $k_0=\pi$ to $k_1=0.28\pi$, supporting its nonlocal and collective nature. Because of this, $f_S$ may not be a good measure of the overall itinerancy of the spins. As may be seen in Fig.~\ref{Fig:f}(a), as $S^{ff}$ becomes saturated at $J_K/t>2.6$, $f_A$ approaches unity accordingly whereas the average local hybridization, $f_S$, continues to increase at much larger $J_K$. The collective nature of the hybridization is best seen at small $n^c$, where one always finds a large Fermi surface at finite $J_K$ despite of the huge deficiency of conduction electrons for screening the local moments.

We now focus on the behavior of $f_A$ and ask if it is a good account of the collective hybridization and what information it may provide us about the evolution of the hybridized system. In the two-fluid model, the hybridization parameter, $f_h$, measures the fraction of the itinerant heavy component of the $f$-electrons and plays a central role in determining the properties of the ground state and the temperature evolution of all thermodynamic and magnetic quantities \cite{Yang2012}. For $f_h<1$, a fraction of the $f$-electrons could stay localized all the way down to zero temperature and form long-range magnetic order; while for $f_h=1$, all $f$-electrons become itinerant and the ground state could be a Fermi liquid. At the delocalization/magnetic quantum critical point, the Fermi surfaces of the conduction electrons may change abruptly to incorporate the $f$-electrons and superconductivity may emerge. If $f_A$ is a good approximation of the hybridization parameter, $f_h$, it should also be correlated with antiferromagnetism and the Fermi surface change \cite{Yang2012}. 

Since antiferromagnetic long-range oder is suppressed in the 1D Kondo-Heisenberg model due to strong quantum fluctuations, we study here the antiferromagnetic correlation energy,
\begin{equation}
E_{H}(J_{K})=J_{H}\sum_{n}\langle \vec{S}_{n}\cdot\vec{S}_{n+1}\rangle.
\end{equation}
We find $E_H(J_K)$ also saturates for $J_K/t\ge4$. For comparison, we take $J_K/t=6$ as a reference for the nonmagnetic ground state and define
\begin{equation}
f_E(J_K)=\frac{E_{H}(J_{K})-E_{H}(0)}{E_{H}(J_K/t=6)-E_{H}(0)}.
\end{equation}
Fig.~\ref{Fig:f}(b) compares $f_E$ with $f_A$ for different values of $n^c$ and $J_H$. We see in all cases, the two quantities are in good agreement; this proves that $f_A$ indeed describes the suppression of antiferromagnetic correlations in the spin chain with increasing hybridization. This conclusion has been further verified in the presence of frustration with next-nearest-neighbor coupling in the spin chain.

Abrupt Fermi surface change has been observed at the quantum critical point of local moment antiferromagnetic order in heavy electron materials \cite{Shishido2006,Friedemann2010}. But in 1D, the exact location of the transition is controversial. While some theory has suggested a Fermi surface change at $J_K=0$ \cite{Yamanaka1997}, DMRG calculations always point to a transition at finite $J_K$ \cite{Moukouri1996,Eidelstein2011}. We will not try to solve this issue here, but only point out that our results show strong correlations between $f_A$ and the redistribution of the conduction electrons in the Brillouin zone. As may be seen in Fig.~\ref{Fig:f}(a), the "critical" coupling $J_K/t=2.8$ at which $f_A$ approaches unity coincides with the coupling at which the maximal slope in $n_k$ is suppressed to nearly zero and suddenly changes its location from $k_F^S$ to $k_F^L$ shown in Fig.~\ref{Fig:nk}(b).

\begin{figure}[t]
\begin{center}
\includegraphics[width=0.4\textwidth]{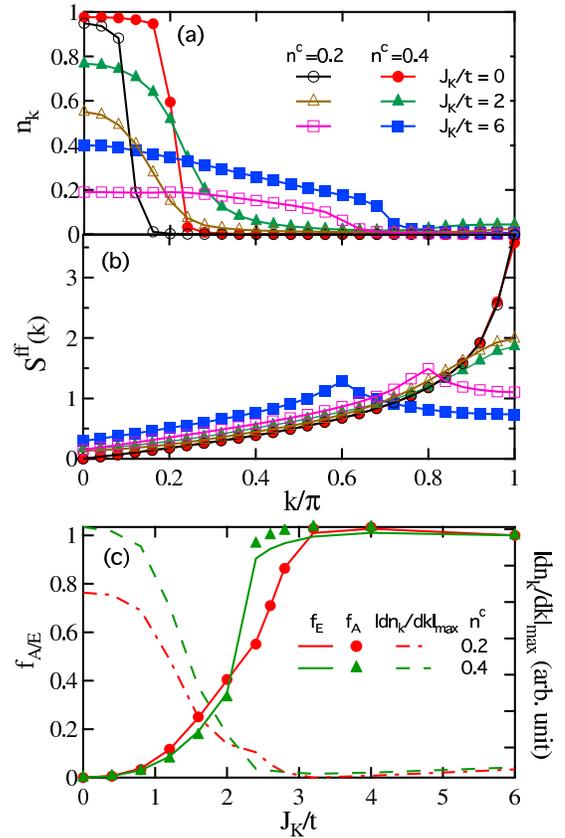}
\caption{(color online) (a) The momentum distribution, $n_k$; (b) the correlation spectrum of the local spins, $S^{ff}(k)$; (c) the hybridization parameters $f_A$ and $f_E$ and the maximal $|dn_{k}/dk|$ as a function of $J_K/t$, for $J_{H}/t=0.5$, $N=50$ and $n^{c}=0.2$ and 0.4.}
\label{Fig:nc}
\end{center}
\end{figure}

The change to a larger volume of the conduction electron Fermi surface with increasing $J_K$ also appears for very small $n^c$, despite of the lack of sufficient conduction electrons to screen the local spins, as shown in Figs.~\ref{Fig:nc}(a) and \ref{Fig:nc}(b) for $n^c=0.2$ and 0.4 at $J_K/t=0, 2$ and 6. The derived hybridization parameters, $f_{A/E}$, are plotted in Fig.~\ref{Fig:nc}(c).  The two are found in reasonable agreement and approach unity at the "critical" $J_K/t$ where the Fermi surface change coincides with the magnetic critical point. These results suggest that the heavy electrons emerging with the large Fermi surface must be of composite nature involving both conduction electrons and magnetic spin fluctuations. The Kondo lattice physics is indeed governed by collective hybridization beyond the local Kondo picture.

The above comparisons confirm that $f_A$ is a good approximation of the hybridization parameter and has indeed the desired properties proposed in the two-fluid model. Because satisfactory theoretical and numerical calculations have not been generally available for quantitative interpretation of experiment, the two-fluid model has been quite successful in providing an intuitive and phenomenological way of organizing the vast amount of complicated experimental data. Our work here provides a plausible justification and the first microscopic explanation for this simple two-fluid procedure of experimental analysis. The observed enhancement of collective hybridization by antiferromagnetic spin correlations at small $J_K/t$ is in distinct contrast to the prevailing wisdom based on their competition and may be the missing piece for a better microscopic theory. Signatures of this enhancement include momentum-dependent delocalization/localization that may be observed using the scanning tunneling or angle-resolved photoemission spectroscopies \cite{Aynajian2012,Mo2012}.

We should note that the mixture of the two components in the spin-correlation spectrum in the momentum space prevents an exact determination of their spectral weight and the approximation might get even worse near half-filling [see Fig.~\ref{Fig:f}(b) for $n^c=0.92$]. Also, in the diluted case, where most local spins are removed, antiferromagnetic spin correlations are suppressed by doping so the proposed separation scheme naturally fails. Nevertheless, we believe it is a good starting point and clarifies the microscopic origin of the "partial" itinerancy and the dual behavior of the $f$-electrons. Our results show that this duality originates from the dynamic hybridization and partial entanglements of the local moments with conduction electrons and should not be confused with the heavy electron condensation in the simplest version of large-$N$ mean-field approximation, although the latter does provide some insights under certain circumstances. We emphasize that the emergence of heavy electrons is  not a phase transition, consistent with the Mermin-Wagner theorem \cite{Mermin1966}. Our proposal may therefore be extended to more realistic situations at higher dimensions and finite temperatures, which are not dealt with in this work due to numerical difficulties. A better separation may even be possible if the dynamic spin-correlation functions are taken into consideration. Recent neutron scattering experiments have observed similar two-fluid behavior in pnictide compounds and found that the local and itinerant components are well separated in energy \cite{Dai2012}. We propose direct probe of the two coexisting fluids in heavy electron materials by neutron scattering measurements. Calculations of the dynamic spin-correlation function are beyond the scope of this study and will be investigated in future work.

To summarize, we use DMRG to study the interplay of the localized and itinerant behaviors in the 1D Kondo-Heisenberg model and find that the dual behavior of the local spins originates from their simultaneous entanglement with other spins in the chain and the conduction electrons in the intermediate coupling regime, as manifested in the distinct peak structures in their spin-correlation spectrum. We propose that the spectral weight of these emergent structures provides a measure of the hybridization parameter in the phenomenological two-fluid description and confirm that it is correlated with the suppression of antiferromagnetism and the change of the Fermi surface. Our results provide a microscopic understanding of the dual nature of the $f$-electrons and may be used as a guide for future experimental and theoretical explorations of the two-fluid physics in real materials. 

We thank David Pines for his helpful comments. This work is supported by the National 973 Project of China (Grant No. 2015CB921303), the National Natural Science Foundation of China (NSFC Grant No. 11174339) and the Strategic Priority Research Program (B) of the Chinese Academy of Sciences (Grant No. XDB07020200).

\end{document}